# An investigation into the vortex formation in a turbulent fluid with an application in tropical storm generation

By Kaushik Majumdar

Electronics & Communication Sciences Unit, Indian Statistical Institute, 203, B. T. Road, Calcutta – 700108, India

The present work discusses about a possible physical interpretation of the occurrence of turbulence in a dynamic fluid with mathematical modeling and computer simulation. Here turbulence is defined to be a phenomenon of random velocity field in the space-time continuum accompanied by chaotic occurrence of vortices. This interpretation is independent of the Navier-Stokes equations. I have reasoned how individual fluid parcels are disintegrated with increasing Reynold's number (Re) (or increasing velocity or decreasing viscosity or both) leading to creation of smaller parcels with arbitrary speeds in arbitrary directions, which destroys the laminar structure of the fluid flow. I have modeled the occurrence of a vortex as a result of collision among linear fluid jets under certain conditions. Chaotic occurrence of such vortices further randomizes the velocity field. These together lead ultimately to turbulence. I have also shown an application of vortex formation in a dynamic fluid in atmospheric science, where it has been shown how an initial disturbing cyclonic vortex is created by collision between two linear wind jets under certain conditions, which under favorable conditions, may mature into a severe tropical storm. Then a three dimensional mathematical modeling of the vortex (assuming that it is going to become a matured storm) has been proposed with computer simulations. This helps us to understand the mystery of origin of cyclonic and anticyclonic vortices in atmosphere and some of their observed asymmetries.

## 1. Introduction

In the study of turbulent flows – as in other fields of scientific enquiry – the ultimate objective is to obtain tractable quantitative theory or model that can be used to calculate quantities of interest and practical relevance. A century of experience has shown that the 'turbulence problem' to be notoriously difficult, and there are no prospects of a simple analytic theory. As a consequence there is no one best model to describe all aspects of a turbulent flow, but rather there is a range of models that can usefully be applied to the broad range of turbulent flow problems.

So far there is no universally agreed upon definition of turbulence. Neither there is any fixed set of criteria satisfiable by a turbulent flow. However *turbulent flow* (or *turbulence* for short) is a dynamic phenomenon signified at least by the following two criteria:

(i)      The velocity field $U(\mathbf{x}, t)$ is random in a turbulent flow.

(ii)   Vortices occur chaotically in a turbulent flow.

$\mathbf{x} = (x_1, x_2, x_3)$. In section 2 I shall investigate into the process of initiation of a turbulence by breaking up the laminar structure of a fluid flow, which eventually leads to randomness in the velocity field and chaotic occurrence of vortices. A possible mechanism for vortex formation from linear fluid jets has been discussed. Occurrence of vortices further randomizes the velocity field. In section 3 I shall study the role of an atmospheric turbulence behind the genesis of some of the tropical cyclones. Occurrence of vortices of sufficiently strong intensity in the atmosphere, which may become an intense tropical cyclone under favorable conditions is still a mystery (Emanuel (1988), Hakim et al. (2002)). In section 3 I have shown that the same mechanism responsible for occurrence of vortices in a turbulent fluid as discussed in section 2 is also responsible for occurrence of cyclonic (and anticyclonic) vortices in atmosphere. That is, a vortex in atmosphere is generated as collision of two wind jets under certain conditions. This can give tangible explanation for asymmetric population of cyclonic and anticyclonic vortices and some of their structural differences as observed in Hakim et al. (2002).

## 2. Initiation of turbulence

Ruelle and Takens (1971) have modeled turbulence by studying the dynamical system theoretic properties of the corresponding Navier-Stokes equations. According to them if the phase or solution space of the Navier-Stokes equations has a strange attractor then the flow modeled by the equations will be called a turbulent flow. Here *strange attractor* means an attractor, which as a set, is the product of a two-dimensional manifold with a cantor set.

Unfortunately existence of a general solution of the Navier-Stokes equations is not yet known (Cannone and Friedlander (2003)). For a complete description of the problem visit http://www.claymath.org/prizeproblems/navier-stokes.pdf. According to Leray (1934), who gave the first weak solutions to the Navier-Stokes equations, during the onset of turbulence the Navier-Stokes equations break down. Such possibility has not been ruled out either by Ruelle and Takens (1971). Hardy and Pomeau (1977) have argued that Navier-Stokes equations do not hold in its usual form for two dimensional incompressible viscous fluids. The results presented in Ladyzhenskaya (1969) support the belief that it is reasonable to use the Navier-Stokes equations to describe the motions of a viscous fluid in the case of Reynold's numbers which do not exceed certain limits. These results also force us to find other explanations for observed phenomena in real fluids, in particular, for the familiar paradoxes involving viscous fluids. She categorically says that (p. 6), "...it is hardly possible to explain the transition from laminar to turbulent flows within the framework of the classical Navier-Stokes theory." Also Navier-Stokes equations do not hold for gases.

According to Landau (1944) the problem of turbulence could appear in a new light if the process of initiation of turbulence could be examined thoroughly. He tried to explain the unsteadiness of laminar motion by superimposing a small disturbing velocity $U_1$ over the original laminar velocity field $U_0$. Hopf (1948) observed that the more is the viscosity the lesser is the effect of turbulence. As the viscosity tends to infinity the flow becomes perfectly laminar and as the viscosity tends to zero the flow becomes turbulent.

In this section I shall try to explain the occurrence of an irregular velocity field in a dynamic fluid independent of the Navier-Stokes equations. Evolution of a vortex at a point is represented by taking the curl of the velocity field at that point. But here I shall take a reverse approach. First I shall explain the chaotic or random occurrence of vortices in a dynamic fluid with high Re, which induces random fluctuation in the velocity field. The method is very simple and the principle guiding motivation will be the observations made by Landau (1944) and Hopf (1948) as cited in the last paragraph.

### 2.1. *The physical background*

Let us consider a hypothetical incompressible fluid with variable viscosity μ. Suppose we have some means to vary μ in the interval [0, ∞). Consider a laminar flow of that fluid for fixed μ. The Reynold's number of the flow is given by Re = ||U||L/μ, where U is the (uniform) velocity field and ||U|| is the euclidean norm of the velocity vector U. L is the dimension of the flow (in some sense), which does not change. Some other dimensionless quantities representing a fluid flow are Rayleigh number (Ra), Prandtl number (Pr), etc., given by

$$\mathrm{Ra} = \frac{\delta g H d^5}{\upsilon \kappa^2 \rho C_p}$$

and

$$\mathrm{Pr} = \frac{\upsilon}{\kappa},$$

where $\delta$ is the volumetric coefficient of thermal expansion, $g$ is the acceleration due to gravity of the Earth, $H$ is the volumetric heating rate, $d$ is the depth of the convection layer, $\upsilon$ is the kinematic viscosity, $\kappa$ the thermal diffusivity, $\rho$ is the average density of the fluid, and $C_p$ is the specific heat at the constant pressure. High values of Ra or Pr produce similar characteristics as of high Re. How a three dimensional flow behaves under heating from below is described by Carrigan (1982) for high values of Ra and Pr. Note that due to heating ||U|| also increases and so does the Re.

What happens if we let μ → 0? Hopf's observation indicates that the flow should become turbulent. How? Well, let us concentrate on Figure 1. If μ is decreased the friction force of sliding a layer of the liquid over any other is also decreased leading to an increase in U (by an increase in U we mean an increase in ||U||). If the density of the fluid is ρ, the parcel P of Figure 1 experiences a normal pressure ρ||U||² on its left flat surface. This pressure is uniformly distributed to each point on the surface of P. At the fluid layers adjacent or close to the boundary of the bulk flow ||U|| has a lower value and hence the fluid experiences a lower pressure due to the velocity. When μ is sufficiently low, ||U|| is sufficiently high and P is sufficiently big, fluid starts flowing out of P in random directions leading to disintegration of the parcel P. A similar breakdown of laminarity has also been observed by Carrigan (1982) for high Ra. The velocity vector field no longer remains uniform. The fluid starts to lose its laminar character, but it is not turbulent yet.

The dynamic fluid losses its laminar nature with increasing Re (or Ra or Pr) and decreasing µ, which is in agreement with observation. In reality where µ remains fixed but U is subjected to increase, the Re is increased and the flow becomes non-laminar due to the same reason.

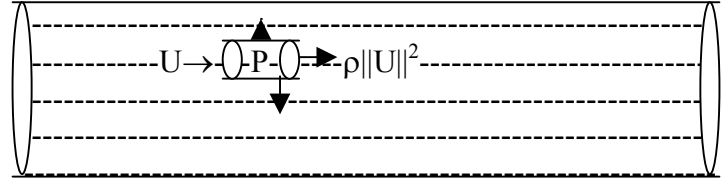

FIGURE 1. A laminar flow of an incompressible fluid, with viscosity µ. P is any cylindrical fluid parcel moving with uniform velocity U. $\rho\|U\|^2$ is the pressure experienced by the parcel due to the velocity U. $\rho$ is the density of the fluid.

### 2.2. *Genesis of vortices*

With decreasing µ and increasing ||U|| the fluid parcels disintegrate as described in the last sub-section. This leads to disintegration of a fluid parcel moving along U into smaller parcels moving in arbitrary directions with random velocities. According to this model, at this stage the vortices are generated. For convenience I have described the process, locally, in polar coordinates. Let (r,θ) be a point where some random fluid parcels are colliding with each other. Let us call colliding fluid parcels as *fluid jets*. In Figure 2 a diagram of collision is presented.

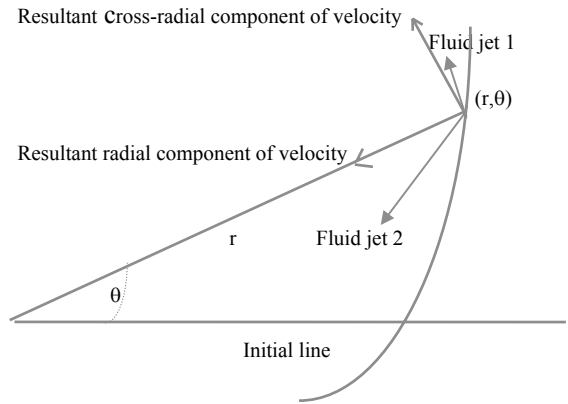

FIGURE 2. Fluid jet 1 and fluid jet 2 are colliding at (r,θ). Their resultant radial component of velocity and resultant cross-radial component of velocity after collision have been shown. If the ratio of these two components is constant a spiral shaped vortex is generated.

From elementary particle dynamics we know that the expression for radial component of velocity at a point (r,θ) is dr/dt and that of the cross-radial component is r(dθ/dt) with respect to some polar coordinate system. If

$$\frac{\text{Resultant radial component of velocity}}{\text{Resultant cross-radial component of velocity}} = \frac{dr/dt}{r(d\theta/dt)} = \pm m(Re), \qquad (1)$$

where m is a function in dimensionless quantity Re. We shall not bother about the exact nature or form of m in this paper, except for the fact that, the values of m are always positive and lie in some suitable compact interval of R – the set of real numbers.

or,  $dr/d\theta = \pm m(Re)r.$  (2)

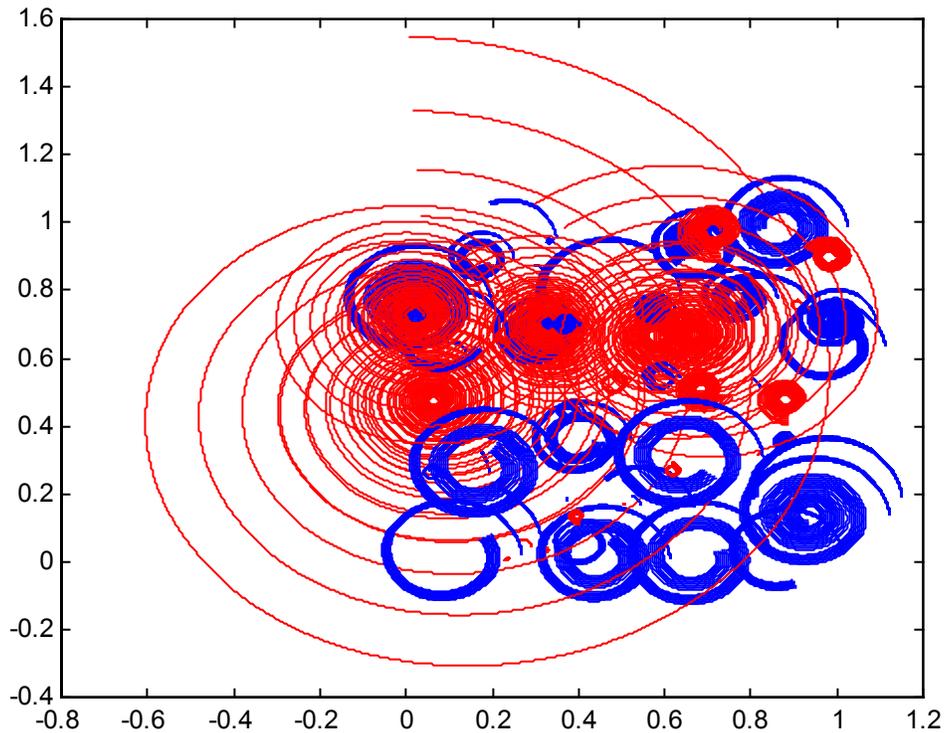

FIGURE 3. Chaotic occurrence of vortices in a two dimensional dynamic fluid. The orientation of the red vortices is anti-clockwise and the orientation of the blue vortices is clockwise. Here [a, b] = [c, d] = [0.05, 0.1]. x = Acos(θ)exp(mθ) and y = Asin(θ)exp(mθ), where A is an integration constant on which the size of the vortices depends. x, y are cartesian coordinates.

In the equations (1) and (2) if m is a constant then the graph of the solution of (2) on the euclidean plane will produce a log-spiral, which describes the shape of a vortex. The sign on the right hand side of (2) determines the orientations of the generated vortex. But in reality m can not be a constant. Instead the values of m are liable to fluctuate within some range. This happens because during the entire collision period momentum transfer takes place between fluid jet 1 and fluid jet 2 of Figure 2 leading to variation in radial and cross-radial components of the velocity field at (r,θ). As a result their ratio ± m(Re) also varies during the entire collision period. m(Re) has another reason to vary also, because during the collision the velocity field at (r,θ) varies, and so also Re varies locally at that point. Without loss of generality we can assume that the values of m(Re) never go beyond certain range, say [a, b], where a, b are real numbers with a ≤ b. So in each collision m actually varies in [c, d], where a ≤ c ≤ d ≤ b. a, b may depend on the nature of the fluid, but c and d vary randomly from point to point. A two dimensional diagram of random occurrence of vortices in a fluid has been presented in Figure 3.

In this hypothetical model only μ is varied and as a result U also varies. In reality μ remains fixed, only U is subjected to change. It is clear that all the above reasoning for evolution of turbulence will remain equally valid if ‖U‖ is increased keeping μ fixed. It is clear from (2) that m(Re) should be such that, in a fluid with high viscosity (that is, the value of μ is very high, which implies, Re → 0) the values of m(Re) will be high in order to make the shape of the vortices more elongated elliptical and therefore will not be able to withstand the shear of the fluid for long and dissipate quickly. In such fluids occurrence of vortices will be much more difficult. We have already discussed that the velocity field in a fluid with high viscosity is difficult to increase. Even if a change is introduced in the velocity field by adding momentum from outside the extra kinetic energy is quickly dissipated by the resistance due to high viscosity.

We have seen that increasing ‖U‖ sufficiently destroys the laminar structure of a fluid flow. Chaotic occurrence of vortices makes the velocity field U even more random. These two together make the flow turbulent.

Some experimental support for this model is not difficult to observe. For example, consider the laminar upward flow of smoke from a lighted incense stick in stationary air (except for the flow due to the thermal convection produced by the lighted incense stick). Perturb the laminar part of the smoke by blowing through a thin nozzle across it. Spiral shaped irregular smoke patterns will form immediately in place of the uniformly linear upward trail of smoke. Similar phenomenon in liquid can also be observed by fitting two pipes, perpendicular to each other, in a large open shallow reservoir and pumping water through both of them simultaneously, but at different velocities. Shape and size of the vortices are to be observed for various ratios of the velocities at the two inlets. Friedlander and Yudovich (1999) have reviewed a number of interesting experiments in their historical perspective with excellent photographs.

This modeling enables us to write the velocity vector field as

$$u_1(x_1, x_2, x_3, t) = v_1(x_1, x_2, x_3, t) + A_1 e^{-m_1 f_1(x_1, x_2, x_3, t)}, \tag{3}$$

$$u_2(x_1, x_2, x_3, t) = v_2(x_1, x_2, x_3, t) + A_2 e^{-m_2 f_2(x_1, x_2, x_3, t)}, \tag{4}$$

$$u_3(x_1, x_2, x_3, t) = v_3(x_1, x_2, x_3, t) + A_3 e^{-m_3 f_3(x_1, x_2, x_3, t)}, \tag{5}$$

where U = $(u_1, u_2, u_3)$, $v_i$ is the linear part of the velocity field and $f_i$ gives the angular displacement in the spiral vortex for all i ∈ {1, 2, 3}. Of course both are scalar fields. $A_i$ gives the size of the vortex, $m_i$ gives the shape of the vortex. $v_i$ varies from $v_j$, for i ≠ j, only in values of some constant terms. Similarly, $f_i$ differs from $f_j$ only in the values of some constant terms involved in the description of the functions. Like $m_i$ the values of $A_i$ too will lie in some compact interval of R for each i ∈ {1, 2, 3}.

## 3. Tropical storms

In the last section initiation of turbulence in a dynamic fluid has been discussed with the help of a hypothetical modeling and simulation. It has long been suspected that the atmospheric turbulence plays a definite role in generating intense tropical storms. In this section it will be argued how the theory developed in the last section helps to confirm this hypothesis.

### 3.1. *Initial conditions for cyclone genesis*

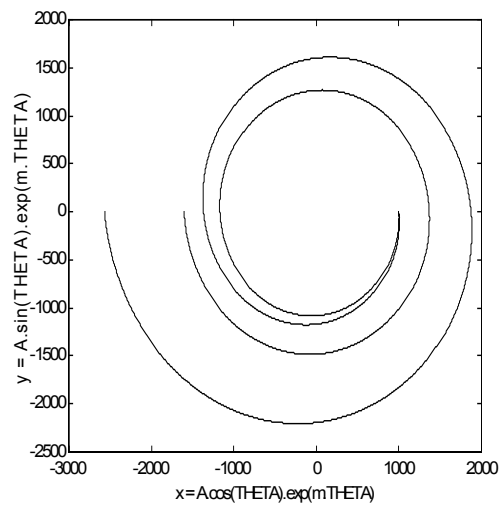

FIGURE 4. Numerical simulation of the lateral cross section of an initial disturbing vortex roughly 1 km above the sea surface in 0.5 km scale. In polar coordinate the diagram will be in the *r-θ* plane, where $r = \sqrt{x^2 + y^2}$ and $\theta = \tan^{-1}(y/x)$. This has originated as a chaotically occurred vortex in a turbulent atmosphere. Comparison of this vortex with the red vortices of Figure 3 will make clear its origin in turbulence.

Geostrophic turbulence has been discussed among others by McWilliams et al. (1994), (1999) and Reinaud et al. (2003). Fluid motions in planetary atmospheres and oceans are often strongly influenced by planetary rotation and stable vertical density stratification. These flows exhibit a rich phenomenology of global-scale circulation and long-lived vortices. Examples of such coherent vortices include Jupiter's red spot, tropospheric cyclones and hurricanes, stratospheric polar vortices, oceanic gulf stream rings, etc.

Weather prediction experts are aware of the existence of several empirical conditions that are necessary but not sufficient for the formation of cyclones (Emanuel (1986), (1988)). The first of these requires the sea temperature to be at least $26^0C$ through a depth of at least 60m. A second requirement is the absence of significant vector changes of the mean wind velocity that extends into the troposphere. Relative humidity will have to be 85% or more for a long time throughout the region of storm formation and development. There are other empirical conditions as well, but even when they are satisfied, storm formation usually does not take place. In fact necessary climatic and geographical conditions for the formation of the tropical storms prevail over large parts of the Earth during the storm seasons, yet the actual occurrence of a tropical storm of sufficient intensity (with a wind speed of 90 km/h or above, to be called a cyclone, hurricane or typhoon) is a relatively rare phenomenon. For the generation of an intense tropical storm creation of a sufficiently strong initial disturbing vortex is essential. Such a vortex is created by a geostrophic turbulence.

According to the study of Emanuel (1988), to get a cyclone started, a vortex, which decays upward from the surface, had to be superimposed on the basic state of formation of the storm. As described in the last section, in atmospheric turbulence too the vortices are generated by linear wind jets (Majumdar (2003)). Two linear wind jets are needed to produce a sufficiently strong vortex. One of them is strong (speed over 40km/h) and the other is weak (speed about 5km/h). The first one acts as the cross-radial velocity and the last one as the radial velocity of the generated vortex. If the vortex is cyclonic it is intensified with the coriolis force. Similarly, if the orientation of the vortex is just the opposite, it is called anti-cyclonic vortex and in that case the coriolis force acts as a damping force to weaken it. Reinaud et al. (2003) have shown that the statistical mean height-to-width aspect ratio of the geostrophic vortices is 0.8. Also two or more vortices of a turbulence may interact leading to dissipation and/or intensification of some of the vortices. It is interesting to note that the existence of a cyclonic vortex has also been detected in the hot liquid outer core inside the Earth (Olson (1999)). The mechanism behind generation of this vortex too may be similar to that for the generation of a tropical storm.

In the storm seasons on a tropical sea under the favorable conditions, as described in the second paragraph of this sub-section, an initial disturbing vortex of right intensity and right size generated due to atmospheric turbulence may mature into a severe cyclone (hurricane in USA). Dynamics of such a vortex is governed by a set of fluid dynamic as well as thermodynamic equations described in the following sub-section.

### 3.2. *Structure of the storm over the PBL*

Here by PBL we mean the planetary boundary layer roughly 1 km above the sea surface. Since the successful modeling of Ooyama (1969), steady state axisymmetric models seem to capture well the basic dynamics of a tropical cyclone. Assume that a sufficiently strong initial disturbing cyclonic vortex has already been generated as a spontaneous evolution of a turbulent vortex in the atmosphere over a tropical sea as described earlier in this paper. It must satisfy the following equation for the conservation of angular momentum.

$$M = rV + \tfrac{1}{2} fr^2, \tag{6}$$

where $M$ is the angular momentum/unit mass, $r$ is the distance between the center and the outermost layer of the log-spiral vortex, $V$ is the tangential or cross-radial component of velocity, and $f$ (assumed constant) is a parameter due to the coriolis force of Earth's rotation about its own axis.

Assuming that the non-adiabatic gas laws are valid we get,

$$\alpha \frac{\partial p}{\partial z} = -g, \qquad (7)$$

$$\alpha \frac{\partial p}{\partial r} = \frac{V^2}{r} + fV = \frac{M^2}{r^3} - \frac{1}{4} f^2 r, \qquad (8)$$

where g is acceleration due to gravity, $p$ is the pressure and $\alpha$ is the volume. $V$ is replaced by equation (6). From (7) and (8) we get,

$$g\left(\frac{\partial z}{\partial r}\right)_p = \frac{M^2}{r^3} - \frac{1}{4} f^2 r. \qquad (9)$$

(7) can be written as

$$g\left(\frac{\partial z}{\partial p}\right)_r = -\alpha. \qquad (10)$$

Differentiating (9) with respect to (w.r.t) $p$ and differentiating (10) w.r.t $r$ and equating $\frac{\partial^2 z}{\partial r \partial p} = \frac{\partial^2 z}{\partial p \partial r}$ we get,

$$\frac{1}{r^3}\left(\frac{\partial M^2}{\partial p}\right)_r = -\left(\frac{\partial \alpha}{\partial r}\right)_p. \qquad (11)$$

Assuming $\alpha$ as a function in $p$ and moist entropy $s^*$ alone and keeping $p$ fixed we get,

$$\frac{\partial \alpha}{\partial r} = \frac{\partial \alpha}{\partial s^*} \frac{\partial s^*}{\partial r}. \qquad (12)$$

From the first law of thermodynamics we get,

$$Tds^* = du + pd\alpha - Ldq^*, \qquad (13)$$

where $u$ is the internal energy, $L$ is the latent heat of vaporization, $q^*$ is the saturation mixing ratio of water vapor with the atmosphere. The saturation moist enthalpy $h^*$ is defined as

$$h^* = u + p\alpha - Lq^*. \qquad (14)$$

From (13) and (14) it follows that,

$$dh^* = Tds^* + \alpha dp. \tag{15}$$

Keeping $s^*$ and $p$ constant respectively the following relations follow from (15).

$$\left(\frac{\partial h^*}{\partial p}\right)_{s^*} = \alpha. \tag{16}$$

$$\left(\frac{\partial h^*}{\partial s^*}\right)_{p} = T. \tag{17}$$

$q^*$ is a function of $T$ and $p$ alone. From (15) it is clear that $h^*$ is a function in $s^*$ and $p$. So we can write (assuming that the $h^*$ is continuously doubly differentiable with respect to both $s^*$ and $p$ irrespective of the order).

$$\left(\frac{\partial}{\partial s^*}\right)_{p}\left(\frac{\partial h}{\partial p}\right)_{s^*} = \left(\frac{\partial}{\partial p}\right)_{s^*}\left(\frac{\partial h^*}{\partial s^*}\right)_{p}. \tag{18}$$

Substituting (16) and (17) in (18) we get,

$$\left(\frac{\partial \alpha}{\partial s^*}\right)_{p} = \left(\frac{\partial T}{\partial p}\right)_{s^*}. \tag{19}$$

Substituting (19) into (12) and then using that in (11) we get,

$$\frac{1}{r^3}\left(\frac{\partial M^2}{\partial p}\right)_{r} = -\left(\frac{\partial T}{\partial p}\right)_{s^*}\left(\frac{\partial s^*}{\partial r}\right)_{p}. \tag{20}$$

We have tacitly assumed that the boundary layer parcels are neutrally buoyant along the angular momentum (*M*) surfaces. This is equivalent to the assumption that the moist entropy of lifted parcels equals the saturated entropy of the environment. This implies that the saturated moist entropy does not vary along the *M* surfaces. Since the moist entropy $s^*$ is then a function of *M* alone we can rewrite (20) as

$$\frac{1}{r^3}\left(\frac{\partial M^2}{\partial p}\right)_{r} = -\left(\frac{\partial T}{\partial p}\right)_{s^*}\frac{ds^*}{dM}\left(\frac{\partial M}{\partial r}\right)_{p}. \tag{21}$$

Dividing both sides of (21) by $\frac{\partial M}{\partial r}$ we get,

$$\left(\frac{\partial r}{\partial p}\right)_M = \frac{r^3}{2M}\frac{ds^*}{dM}\left(\frac{\partial T}{\partial p}\right)_{s^*}, \qquad (22)$$

which is an expression for the $M$ surfaces. Integrating (22) we get,

$$r^{-2} = -\frac{\frac{ds^*}{dM}[T - T_{out}]}{M}, \qquad (23)$$

where $T_{out}$ is an integration constant, which may be a function of $M$ or $s^*$.

$$r^2[T - T_{out}] = -\frac{M}{\frac{ds^*}{dM}}. \qquad (24)$$

$s^*$ is a function in $M$ alone. So from (24) we get,

$$r^2[T - T_{out}]s^* = -\frac{1}{2}M^2 + C. \qquad (25)$$

$C$ is an integration constant. If $T = T_{out}$ then $C = \frac{1}{2}M_0^2$, where $M_0$ is the value of $M$ at $T = T_{out}$. Let us decide a temperature scale in such a way that $T_{out} = 0$. As height increases $T$ decreases. So $T$ is always negative. Then (25) becomes

$$r^2 T = \frac{M_0^2 - M^2}{2s^*}. \qquad (26)$$

From non-adiabatic gas laws we get,

$$-\rho g z^2 = RT, \qquad (27)$$

$\rho$ is density of the atmosphere and $R$ is gas constant. Eliminating $T$ from (26) and (27) we get,

$$r^2 z^2 = \frac{R(M^2 - M_0^2)}{2\rho g s^*}. \qquad (28)$$

Clearly, for a given value of $M$ $s^*$ is also fixed (for $s^*$ is a function in $M$ alone). In a buoyant air parcel $\rho$ is also fixed. Under these conditions (28) becomes

$$rz = \pm\sqrt{\frac{R(M^2 - M_0^2)}{2\rho g s^*}} \ . \tag{29}$$

In (29) ± sign is taken to signify the change of orientation of the rectangular hyperbola in the *r-z* plane. When $T > T_{out}$ + sign is taken and when $T < T_{out}$ the – sign is taken. Here we have taken $T_{out}$ to be the temperature roughly 1 km above the sea surface. $T < T_{out}$ above that level and therefore we shall use the negative sign only in our simulations (Figure 5).

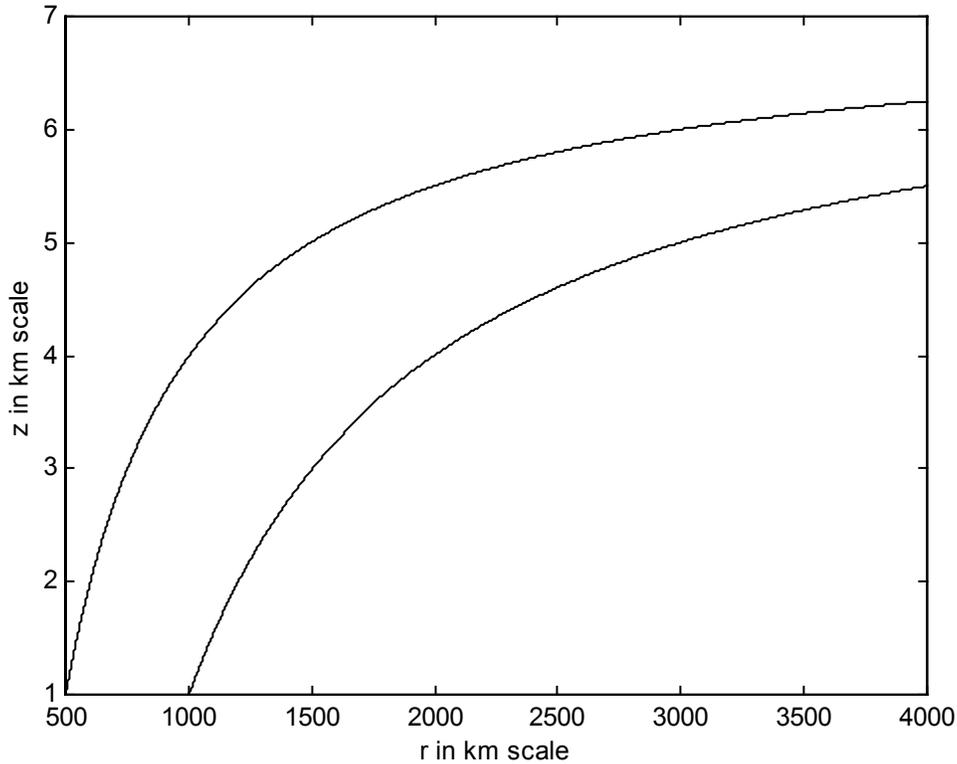

FIGURE 5. Constant momentum surfaces in *r-z* plane. These surfaces give the boundary walls of a cyclonic vortex. In other words this is the longitudinal cross section of a three dimensional cyclonic vortex.

Figure 4 and Figure 5 together give the cylindrical structure (rather like a trough) in a polar coordinate system (*r, θ, z*) of the formation and development of a cyclone above a tropical sea. According to Emanuel (1987) tropical cyclones make text book examples of Carnot engine, with the provision that the heat input is largely in the form of the latent heat of vaporization. Figure 5 illustrates the energy cycle. At some radial distance from the storm center, assumed at (0,0) in Figure 4, typically 100 to 500 km, surface air begins to flow inwards towards the storm center within a frictional boundary layer whose depth is about 1 to 2 km. During its inward trek the air maintains a nearly constant temperature that is very close to the sea surface temperature but acquires water vapor from the sea, which supplies the latent heat of vaporization. The rate at which it acquires latent heat is a monotonic function of near surface wind speed. As the air flows toward lower pressure, heat is added due to isothermal expansion as well. It is also during this inflow that the air suffers the greatest frictional dissipation. At a much smaller radial distance (from the storm center), typically from 5 to 100 km (the simulation in Figure 5 is at a much larger scale, for it gives only the starting stage of the storm), the air abruptly turns upward and ascends through the deep cumulonimbus cloud that constitutes the eye wall. During this ascent the total heat content (sensible plus latent) is approximately conserved, though in the process there are large conversions of latent to sensible heat. There is also comparatively little frictional loss of energy in this branch. Finally, the air flows outward at the top of the storm and eventually loses the heat gained from the sea by long-wave radiation to the space, and through friction reacquires the lost angular momentum. This usually happens at some radial distance which is much larger than the previous one (that is, 100 to 500 km from the storm center). Calculations are rather insensitive to the exact values of the radial distances. The whole process is considered to take place in an atmosphere which is neutral to buoyant and centrifugal convection along angular momentum surfaces.

**4. Conclusion**

I started this paper by proposing a physical interpretation of turbulence in a dynamic fluid, which does not depend on the Navier-Stokes equations. Here turbulence has been viewed as a phenomenon of a dynamic fluid, where the velocity field is random and chaotic occurrence of vortices takes place. In this paper I have argued that a vortex is created due to collision among linear fluid jets under certain conditions. The same interpretation can be carried out to gases also. This has been utilized to show a close relationship between atmospheric turbulence and genesis of tropical cyclones, where the initial disturbing vortex is hypothesized to have been created due to the collision of two linear wind jets, one is weak and the other is strong, under certain conditions. I have given the complete three dimensional modeling and simulation of the cyclonic vortex in a cylindrical coordinate system. The theory of genesis and development of a cyclone is mainly based on the fundamental works of Emanuel (1986), (1987) and (1988).

Asymmetries between cyclonic and anti-cyclonic vortices in terms of both population number and structure have been observed by Hakim et al. (2002). Cyclonic vortices on the tropopause are characterized by compact structure and larger pressure, wind and temperature perturbations when compared to broader weaker anticyclones. To some extent these observations can be explained by the theories developed in this paper. In Figure 3 it has been demonstrated how clockwise and anti-clockwise vortices are randomly generated in a two dimensional turbulent fluid. It is clear from Figure 3 that the anti-clockwise vortices are converging towards the center and the clockwise vortices are diverging away from the center. When cyclonic (anti-clockwise in northern hemisphere) and anti-cyclonic (clockwise in northern hemisphere) vortices are created due to atmospheric turbulence they too have asymmetry in population and structure for the similar reasons. Inward converging nature of cyclonic vortices make them cluster around to form greater and more powerful vortices, whereas anti-cyclones do not due to their outward repelling nature. Similar reasoning with slight modification will hold for the southern hemisphere also (Majumdar (2003)). "Cyclones tend to have plateau like structure, whereas anti-cyclones tend to have broad, sprawling structure" (Hakim et al. (2002)). This observation too is in conformity with the explanation given above.